# Background

Charge has long been thought a real substance. Supposedly contained within or on the surface of a small spherical volume, it even defines the size of a charged elementary particle. If charge occupies space, charge density is meaningful. Maxwell's equations say we can find charge density from divergence **E**. What is charge? Charge is a total mystery. Charge cannot be "explained". It just "is". We know of only one unit of charge, e, amongst all free elementary charged particles.

About 1831, Faraday thought on the nature of the Coulomb force and dismissed "force at a distance" in favor of an invisible static field he named the electric field **E**. He believed this **E** field to be sourced by charge, a concept that quantifies electrification. He did not concern himself with "what is charge", but merely accepted its existence.

From the Lorentz force law, we recognize that both this **E** field and motion relative to a magnetic **B** field exert force on a test charge.

**F** = q(**E** + **vxB**)     [1]

This defines **E** and **B**, in terms of the force exerted on a test charge, q. In many ways, **vxB** is like an electric field. While it is simple and convenient to attribute **E** to the presence of charge, problems arise when we adhere to the concept of charge as a real substance. The first problem is, why does the particle not explode under Coulombic repulsion? A second problem is the size of a charged elementary particle. If the size of an electron were a measure of the volume containing the charge, then elastic Coulomb scattering should be able to find the size. With accelerators, colliding electrons can be given enough energy to partially overcome Coulomb repulsion and cause an overlap of the spherical regions containing charge. This should diminish the Coulomb repulsion, change the scattering pattern, and so let us measure size. It doesn't work. Experimentally, the electron shows itself as a point particle without size. This is devastating, because the energy content of the electric field is infinite for a point particle. A third problem is that the electron carries angular momentum but a point particle cannot. The problems associated with assigning charge as the source of the electric field continue to increase, the more we learn about electrons.

Pair production adds to these problems. If there really were this thing we call charge, it could only be made from energetic EM fields. A few years ago, I noticed that a spinning dipolar **B** field creates, by **vxB**, an electric field that leads to a net electric flux through a surrounding Gaussian surface. This implies charge creation. In effect, I discovered the homopolar generator some 150 years after Faraday. A measurable d.c. voltage can be found by spinning a cylindrical bar magnet about its symmetry axis, with the leads of the voltmeter rubbing against the pole and the waist of the magnet. What is the seat of the emf? No one has yet proved whether the **B** field from a bar magnet is itself spinning or whether the voltage measured lies with metal cutting through a stationary **B** field. Faraday decided the **B** field was not spinning. His contemporary Weber suggested the **B** field was spinning, which would create at every point in space a **vxB** electric field. This conundrum does not arise with the electron. An electron has a permanent magnetic moment and carries angular momentum, and it follows that the **B** field is indeed spinning. This seemed promising for charge creation, but still lacked a plausible mechanism that tied together EM fields stably in an elementary particle such as an electron.

# A model for creating a PEP electron from EM fields only

A simple model of the electron as a PEP (purely electromagnetic particle) has been found that exhibits all the particle properties of stability, charge, rest mass, spin, angular momentum and magnetic moment. (1) This model does not presuppose the existence of "charge", and is built entirely from EM fields. It then finds "charge" to be a mathematical construct. EM fields oscillate at the Compton frequency ($\nu_C = mc^2/h$) between configurations as a magnetic dipole $\mu$ and an electric toroid. The magnetic dipole carries one quantum of flux, h/2e (or maybe h/4e, discussed later). Stability is dynamic, with energy and angular momentum alternately carried by one or the other or some combination of these two configurations. Spin ensures that the **B** field rotates and produces, by **vxB**, a field very like the Coulombic electric field we attribute to charge. This **vxB** electric field is inverse square at all "r", since relativity corrections to "r" and "v" cancel. The rest mass is not predicted, but is recognized as the total internal kinetic energy, divided by $c^2$. The model accounts for charge, e.

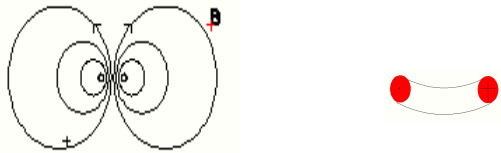

Figure 1. Model of the PEP, alternately configured as a magnetic dipole and electric toroid.

By Gauss' law, "charge" has been created if the integral of **vxB** over a closed surface does not vanish. The **B** field spins about the magnetic dipolar axis. A word of caution. If oscillation of the magnetic dipole reverses polarity, the average "charge" is zero. Consider a magnetic dipole generated by a current, as in a loop of wire carrying current. When the current is interrupted, i.e. when the source of the dipole is removed, the collapsing **B** field creates an electric field that tends to perpetuate the magnetic field. The **B** field energy is then dumped, mostly as a spark when the current switch is opened. When a capacitor is connected across the switch before opening it, there need be no spark and oscillation begins as shown next. Energy is exchanged between **E** and **B** fields. The **E** field is shaped like a toroid, $90^0$ out of phase with **B**. The total energy of the LC oscillating circuit goes as $\mathbf{B}^2$ plus $\mathbf{E}^2$, and damps out over time due to ohmic losses and EM radiation. These thoughts guide the model.

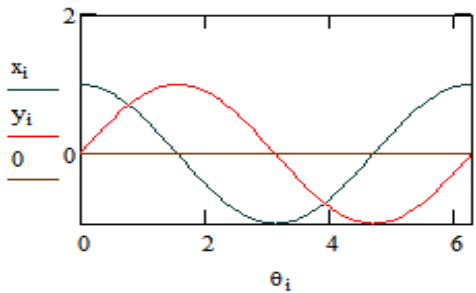

Figure 2. LC circuit oscillating, **B** is black and **E** is red.

The electron has a permanent magnetic moment $\mu$. We need to find a way to incorporated this into the model of an electron. One way of doing this is by postulating that the electron entraps a single magnetic flux quantum. Oscillation then leads to **B** fields that undulate but do not reverse polarity. The magnetic moment undulates. In other words, the magnetic dipole moment has two parts. One comes from a conserved magnetic flux quantum and the other from a transient magnetic moment that oscillates about zero. When an isolated magnetic dipole begins to decay, it gives rise to a toroidal electric field that carries Maxwell's displacement current, $\varepsilon_0 \, \partial \mathbf{E}/\partial t$, in such a direction as to perpetuate the **B** field. This is illustrated in Figure 3. Notice that $\partial \mathbf{E}/\partial t$ is a cosine function like **B**, and tracks with **B**. The dipolar **B** field is however offset by a constant amount, due to the incorporation of a quantum of magnetic flux.

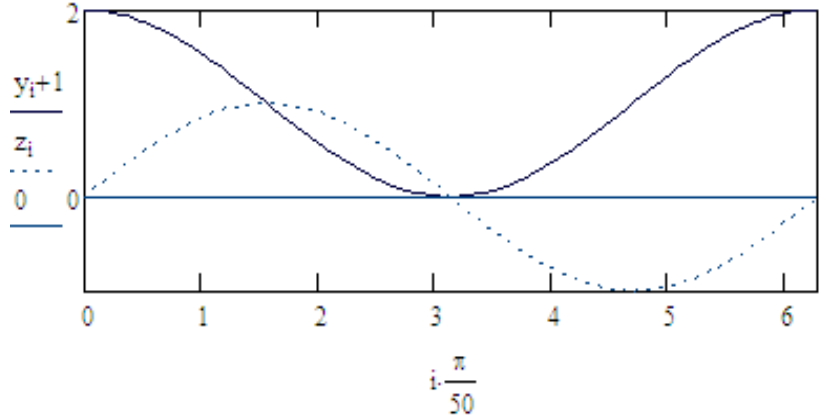

Figure 3. Solid line is **B** and dotted line is **E**.

Consider a magnetic dipole, $\mu$, isolated in space. In atomic spectroscopy, we normally think of the spin angular momentum of an electron as S=½ and of the total angular momentum as L=√(S(S+1))=(√3)/2, in units of h/2π. In this model of an electron, $\mu$ carries spin **L** and spins at a rate we will later associate with √3 times the Compton frequency, $\nu_C$=mc²/h. $\mu$ also undulates, without reversing polarity, at a rate we assume to be the Compton frequency. Does this spin orientation in space remain fixed? If this were a mechanical gyroscope, identifying spin **L** with total angular momentum, the answer is "yes". But, this cannot be. The spinning EM field is anisotropic, but its energy can be lowered by averaging the spin over 3 orthogonal directions. This is achieved by precession of the total spin **L** at arccos 1/√3 about **S**. The motion of the EM assembly can be visualized as like a poorly thrown football. The component of spin along the wobble axis is **S**, and spin **L** lies along the long axis of the football. The **vxB** field created by spinning **B** is also anisotropic, largest at the waist. Precession or wobbling makes it, on average, isotropic.

The electron can be pictured as follows. $\mu$ is oriented along **L**, and undulates at $\nu_C$. The vector average of **L** is **S**, and the direction of **S** is fixed in space. The spinning **B** field about $\mu$ creates an electric field **vxB** at every point in space, and Gauss' law finds that a net charge has been created. The measured magnetic moment of an electron is approximately one Bohr magneton, $\mu_B$. The Bohr magneton, eh/4πm, is the magnetic moment generated by circular motion of an electron in the lowest level of a Bohr atom. It is useful because it is approximately the measured magnetic moment, but has no fundamental claim to being the magnetic moment of an electron. The amplitude of $\mu$, deemed $\mu_C$, must exceed $\mu_B$ because $\mu$ oscillates and further is not aligned with **S**. It will be shown later [5] that the "charge" created by this model is proportional to the product of $\nu_C$ and $\mu_C$. If we choose the spin rate as $\nu_C$, then $\mu_C$=6$\mu_B$ is needed to make charge "e". My preference is to choose a spin rate of √3$\nu_C$ and $\mu_C$=2√3$\mu_B$. With a Gaussian surface enclosing a charge Q, the electric flux is

$$\Phi_E = \int \mathbf{v} \times \mathbf{B} \cdot d\mathbf{S} = Q/\varepsilon_0 \tag{2}$$

To calculate Q for a spinning magnetic dipole with a constant magnetic moment $\mu$, we need the tangential velocity and the tangential component of **B**. In the far field,

$$\mathbf{B} = (\mu_0/4\pi)(\mu/r^3)[\mathbf{r_0}\, 2\cos\theta + \boldsymbol{\theta_0}\sin\theta] \quad\text{and so}\quad B_\theta = (\mu_0/4\pi)(\mu/r^3)\sin\theta \tag{3}$$

A **vxB** electric field is created at every point in space by the spinning **B** field. **vxB** remains inverse square at all "r", thanks to compensating SR factors. For a spin rate of $\sqrt{3}\nu_C$, the tangential velocity and the radial component of **vxB** are:

$$v = (2\sqrt{3}\pi mc^2/h)\, r\sin\theta \qquad (\mathbf{v}\times\mathbf{B})_r = (\mu_0/4\pi)(\omega\mu/r^2)(\sin^2\theta)\mathbf{r_0} \tag{4}$$

In this model, the magnetic moment undulates sinusoidally with time, and its average value is half $\mu_C$. The measured magnetic moment is even less, because of precession of $\mu$ about **S** at arccos $1/\sqrt{3}$. This means that $\mu_C$ must be chosen larger than the measured value, $\mu_B$, by $2\sqrt{3}$. This is the reason for choosing the spin rate $\sqrt{3}\nu_C$. The component of spin along **S** is $\nu_C$. Calculation of charge is straightforward.

$$Q = \varepsilon_0\,\Phi_E = \varepsilon_0 \int (\mu_0/4\pi)(2\sqrt{3}\,\mu_B/r^3)\sin\theta(2\sqrt{3}\pi mc^2/h)\, r\sin\theta\, 2\pi\, r^2\sin\theta\, d\theta = e \tag{5}$$

Many people have pondered over why we find in nature only one unit of charge. This model produces exactly the same unit of charge, e, no matter the mass m of the particle. Instead of $\mu_B$, one uses the appropriate value of m in $eh/4\pi m$. This means that every elementary charged particle arising from this mechanism will have exactly the same unit of charge. The sign depends on the direction of spin. Why should charge be independent of mass? A "hand-waving" explanation is that magnetic moment is something like volume ($\lambda_C^3$) times the **B** field within that volume. The size of the core of the magnetic dipole is assumed to be of the order of the Compton wavelength, $\lambda_C = h/mc$. With the magnetic flux quantized, $\mathbf{B}\,\lambda_C^2$ is constant and so the intrinsic magnetic moment $\mu_C$ scales as $\lambda_C$ or as $1/m$. Why do stable charged particles only exist when the mass is that of an electron or a proton? Don't know.

The gyroscopic consequences of precession may be of some consequence. Precession is driven, in this model, by the need to minimize the EM field energy by making the fields, on average, isotropic. Energetically, the precession occurs because **L** is always seeking a lower energy configuration. It is well known that a gyroscope, forced to precess, will exert a torque directed orthogonal to the precession. We have chosen the precession angle as arccos $1/\sqrt{3}$, so as to make the average field isotropic. Perhaps the reaction torque from forced precession changes this angle slightly, in which case the **vxB** field is not wholly isotropic. Of course, the precession rate in this model is unspecified. Any such problems are minimal for a relatively slow precession. An interesting possibility is that this could support an alternate explanation of why the "g" factor for an electron is not exactly 2, and may lead to finding the actual precession rate.

## Stability of the magnetic flux quantum

Ever since I first learned about pair production, I have been fascinated by the way nature ties enduring knots in the EM fields that constitute particles. In general, we expect that any isolated system of EM fields must either decay or fly away at c as a photon. The PEP concept holds that this does not happen, and that the EM fields are somehow tied together to make a stable particle. It seems obvious that the stability of these EM fields must be dynamic, and quantization in some form is needed for stability. The model described here relies on quantized magnetic flux. About 50 years ago, it was proposed (2) and experimentally verified (3) that magnetic flux is quantized. The experimental proof involves the step-wise increase of magnetic moment within a superconducting ring, when forced to accommodate larger and larger currents. Instead of London's predicted value of $\Phi_B$ =h/e, it was found that the magnetic flux increased in units of h/2e. This was "explained" in terms of paired electrons, the current-carrying charge being 2e in a superconductor. In a paper published a decade ago on the APS eprint server (now unfortunately defunct), and which remains available in book form (4), a case is made that the quantum of magnetic flux is h/4e. This alternate assignment of the size of the flux quantum also conforms with the results based on a superconducting ring, the two electrons each supporting a magnetic flux of h/4e. The important thing is that a single quantum of magnetic flux can account for the electron's magnetic moment. But why should this quantum of magnetic flux be so stable, absent the superconducting ring?

Trapped within a superconducting ring, a quantum of magnetic flux is stabilized by a ring current carried by a pair of electrons. Any fluctuation of the flux tends to create an electric field within the ring, and a superconductor cannot support an electric field. The ring current reacts so as to maintain the flux. This means that a flux quantum, surrounded by a superconducting ring, is stable. There is no reason to think this magnetic flux is itself spinning. Consider the current as consisting of a charge circling the ring, at the speed needed to create the current that supports the magnetic flux quantum. Suppose we transform to a rotating coordinate system, rotating at the same speed as the charge. The charge is then stationary, and we have transformed away the current. So, what supports the magnetic flux quantum in this coordinate system? We find the dipolar **B** field is now spinning, in the opposite direction and at the same rate. This suggests that a spinning dipolar magnetic moment is equivalent to a non-spinning flux quantum surrounded by a stabilizing current. It should be noted that a steady-state dipolar **B** field derives from the gradient of a scalar, and so the addition of an entrapped magnetic flux quantum configured as a magnetic dipole is simply a gage transformation.

## Magnetic monopoles?

It is interesting to recognize that the equivalent of a magnetic monopole would be created, were we able to exchange the configurations of the electric and magnetic fields in this charge creation model. **B=vxE**/$c^2$ and it follows that $(1/c^2)\int$**vxE.dS** does not vanish. Gauss' law then finds the presence of magnetic monopoles within a closed surface. Such a magnetic monopole would exhibit a permanent electric dipole, needed to source the spinning **E** fields.

## Where is the charge, in this model?

Gauss' law finds charge in this model, and in differential form even tells us how to find it. Before doing the calculations, lets try to localize any areas that contain charge using the integral form of Gauss' law. Consider the dipolar **B** field.

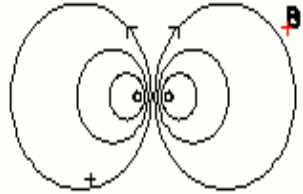

Figure 4. A dipolar **B** field.

Focus on an individual closed loop of **B** in the figure. With the assembly rotating about the vertical symmetry axis, the outer parts of that loop exhibit a **vxB** field pointing inwards. But on the inner parts of the loop, near the spin axis, the **B** field reverses and **vxB** points outwards. Lets construct a Gaussian surface, shaped like a toroid, whose cross-section encompasses the loop of **B** field under consideration. It is clear that the net **vxB** electric flux is negative definite. This suggests the presence of a negative charge density topologically confined, more or less, to a plane perpendicular to the spin axis. There is also an imputed positive charge near the center. This derives from another Gaussian surface, shaped like an apple core and centered on the EM assembly. The sides of the "apple core" Gaussian surface are very near center, such that the **vxB** electric field points outwards. The end caps of this surface are drawn orthogonal to the **B** field, so that no **E** fields enter or leave. Since the net flux is outward, this finds positive charge density distributed more or less along the symmetry axis.

The distribution of "charge density" in this model is of interest, since size is generally accepted as whatever part contains the charge. When charge density is calculated by the differential form of Maxwell's law, by taking the divergence of **vxB**, one is surprised to find that no net charge has been produced!  Because this was unexpected, the derivation is shown below in some detail.  The distribution of the magnetic field, far from the magnetic dipole, is simple.

$\quad$ **B** = $(\mu_0/4\pi)$ $(\mu/r^3)[\mathbf{r_0}\, 2\cos\theta + \boldsymbol{\theta_0} \sin\theta]$ $\hfill$ [6]

The unit vectors in spherical coordinates are $\mathbf{r_0}$, $\boldsymbol{\theta_0}$, and $\boldsymbol{\phi_0}$. The tangential velocity is

$\quad$ **v** = $\omega\, r \sin\theta\, \boldsymbol{\phi_0}$ $\hfill$ [7]

"$\omega$" is the angular frequency. The electric field is

$\quad$ **E** = **vxB** $\hfill$ [8]

Cross products of the unit vectors are:

$\quad$ $\boldsymbol{\phi_0} \times \mathbf{r_0} = -\boldsymbol{\theta_0}$ $\hfill$ [9]

$$\boldsymbol{\phi_0} \times \boldsymbol{\theta_0} = \mathbf{r_0} \qquad [10]$$

The **vxB** electric field that will be queried to find the charge distribution is:

$$\mathbf{vxB} = \{\omega r \sin\theta\, \boldsymbol{\phi_0}\} \times \{(\mu_0/4\pi)(\mu/r^3)[\mathbf{r_0}\, 2\cos\theta + \boldsymbol{\theta_0} \sin\theta]\} \qquad [11]$$

$$\mathbf{vxB} = (\mu_0/4\pi)(\omega\mu/r^2)[-\boldsymbol{\theta_0}\, 2\cos\theta \sin\theta + \mathbf{r_0} \sin^2\theta] \qquad [12]$$

In spherical coordinates, the operator for divergence of a vector **A** is

$$\text{div } \mathbf{A} = (1/r^2)(\partial/\partial r)(r^2 A_r) + (1/r\sin\theta)(\partial/\partial\theta)(A_\theta \sin\theta) + (1/r\sin\theta)(\partial/\partial\phi) A_\phi \qquad [13]$$

The first and last terms are zero. The charge density is then calculated, using Maxwell's equation.

$$\rho = \varepsilon_0 \text{ div } \mathbf{vxB} \qquad [14]$$

The result is

$$\rho = (\mu_0 \varepsilon_0/2\pi)(\omega\mu/r^3)(3\cos^2\theta - 1) \qquad [15]$$

This poses several problems. One expects the integral of charge density over the volume to yield the total enclosed charge, Q. Instead, integration finds Q=0 no matter the radius.

$$Q = \iiint \rho\, dV = \iint (\mu_0 \varepsilon_0/2\pi)(\omega\mu/r^3)(3\cos^2\theta - 1)(2\pi r^2 \sin\theta\, d\theta\, dr) = 0 \qquad [16]$$

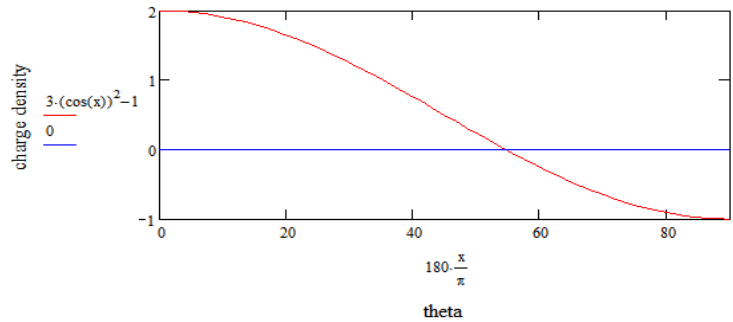

Figure 5. θ dependence of div **vxB**.

Locally, "charge density" per [15] is not zero but the integrated charge density is zero! [15] finds both a positive charge density along the spin axis and a negative charge density along the waist. Of course, this calculation is for the instantaneous anisotropic **vxB** field with fixed orientation of the spin. Averaged over time, **vxB** is isotropic and so <div **vxB**> = 0 at every point. Upon reflection, it is clear why one cannot find "charge density" in this model, on average, simply because <**vxB**> [12] is everywhere inverse square. Any volume between two concentric spheres will intercept the same electric flux, inward and outward.

It gets worse! When the size of an electron is interrogated by elastic scattering of fast electrons, one looks for any failure of the inverse square Coulomb repulsion when the bombarding missile penetrates the purported volume of the electron. From [12], the radial component of **vxB** is everywhere inverse square. This means that an inverse square Coulomb repulsion exists at all r, even as r→0 in this model. If the magnetic dipole has any size at all, this will change [12] at small r. And if scattering finds this size, eventually, the force law will probably be something other than Coulombic. So, how big is an electron? Measured by Coulomb repulsion, it has no size and would appear in the next figure as a vertical line along the "y" axis. Distance "r" is in arbitrary units. Measured by the **B** field, shown in red, the size of the electron is infinitely large. Measured by the **vxB** field, created in place at every point in space and not derived from an isolated charge near the origin, the size is shown in blue and is likewise large. Unless there is a cutoff distance of the Coulomb force between charges, the spinning assembly of the **B** field constituting the electron extends to infinity. Measured by its EM fields, the electron is huge! Measured by Coulomb repulsion, it is a point particle without size.

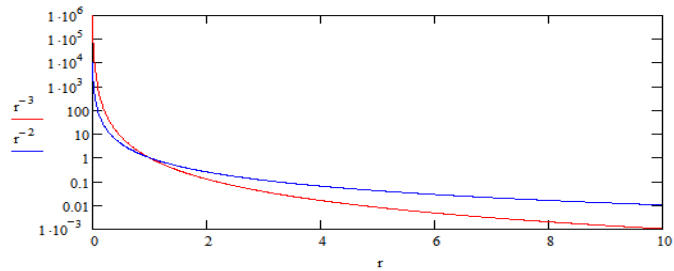

Figure 6. Size of electron? **B** field is red, **vxB** is blue.

## Problems with Gauss' law?

In integral form, Gauss' law anticipates that the only source of electric flux at the surface lies with charges contained within. This model of the electron finds a net **vxB** electric flux at the Gaussian surface, but finds no charge density, on average, within. It is puzzling, and will be examined further in the appendix.

One could object that best efforts have revealed neither size nor structure of an electron when interrogated by fast electrons. The answer lies in part with the relativistic shrinkage of particle size as speed increases. The size of any particle used as a probe shrinks with speed but, in center of mass coordinates, so does the particle being interrogated. Higher energy is ultimately self-defeating. Beyond this, and far more important, is that the Coulombic elastic scattering experiment is wrong-headed. It assumes that all the charge of an electron is confined within a compact volume that defines the electron's size, and seeks to find a change of scattering pattern for those missiles that momentarily cohabit that compact volume. **vxB** generates electric fields similar to those imputed to arise from a compact region of charge, but we try in vain to find any "charge" at center. The range of the **vxB** field presumably extends to infinity, and all attempts to measure the size and structure of the EM fields that constitute an electron by Coulomb scattering are doomed. Space based experiments at large distances might be useful, to test the range of the **vxB** field.

## Deep inelastic scattering

Studies of the structure of protons by scattering very fast electrons have found patterns consistent with the presence of several point particles known as quarks. If the charge of quarks arises from the same mechanism postulated here for an electron, they must exhibit elastic Coulomb scattering as though they were point particles.  Deep inelastic scattering is surely different, in that it deposits energy sufficient to destroy the target.  The ways in which the resulting energetic EM fields rearrange themselves into new particles, on exiting the target area, are of course interesting   But since the target itself has been destroyed, it tells us little about the size and shape and structure of that which was. The conclusion that quarks are point-like does not mean that they are without size or structure.  The very fact of being "charged", by the logic of this study,  requires the existence of a vast assembly of spinning **B** fields. Deep inelastic scattering is somewhat like pair production, where the bombarding photon is broken into parts and these reform as a pair of elementary charged particles.

## Conclusions

Is this merely taxonomy, a renaming of charge in favor of rotating **B** fields as the source of a static electric field?  I think not.  Several things change.  The **vxB** electric field is not static, but undulates at the Compton frequency. Maxwell's equations deal with transient sources plus a static source called charge.  Since the static electric field **E** does not exist, and since charge and charge density are not real, these concepts should be expunged from Maxwell's equations.  Or, if one retains **E** in Maxwell's equations for mathematical convenience, we must recognize that this is a very different electric  field than the one described by Faraday. The presence of an entrapped magnetic flux quantum to the cleansed Maxwell's equations causes no problem and is simply a gage transformation.  **vxB** electric fields can be  either solenoidal or conservative, according to the geometry of the mechanism by which they are produced. This will be discussed in the appendix.  Perhaps most important of all, size of a charged elementary particle is no longer defined by the volume that contains charge.  At very small radii, [12] must eventually fail and one must take into account the size and structure of the magnetic dipole.  When one does so, the scattering theory is likely to change because the interaction may be magnetic rather than Coulombic.

Most advances in physics are "discoveries".  A few are "undiscoveries", as with the luminiferous ether.  Perhaps this undiscovery of the substance called charge and of the static electric field **E** will also be considered an advance. The PEP concept holds that all particles are ultimately composed of EM fields.  A model of elementary charged particles must be simple, since nature builds so many of them.  We may have finally reached the root of all things, energetic EM fields.  As to why there is only one unit of charge found in nature, this model agrees that charge is independent of mass.  Unless there is a different mechanism for charge creation, the best hope for finding an elementary particle with charge other than e is that particles exist or can be made that contain more or less than one unit of quantized magnetic  flux. Or, a different "spin" might lead to a different charge.  It should be noted that photons presumably contain 2 units of  quantized EM flux, but the absence of observable spin precludes that charge be created.  There can be no "charge" without spin.  It has not escaped notice that a similar model, with a transient magnetic dipole and electric toroid each oscillating about zero, would have many of the attributes of a neutrino.

## Appendix: Gauss' law, Maxwell's laws, and the electric field

Electromagnetic theory attempts to explain the relationship between charge and electric and magnetic fields. These concepts arose separately, and through the genius of Faraday and Maxwell were combined and interrelated. We have also learned, from Einstein's insight, that **E** and **B** are interchangeable according to the observer's motion and have no independent existence. In a vacuum, the Lorentz force exerted on a test charge q defines operationally the **E** and **B** fields.

$$\mathbf{F} = q(\mathbf{E}+\mathbf{v} \times \mathbf{B}) \qquad [1A]$$

About 150 years ago, Maxwell collected the known EM relationships, added the displacement current $\varepsilon_0 \partial \mathbf{E}/\partial t$ to curl **B**, and found the almost symmetric equations known and revered as Maxwell's equations.

$$\text{curl } \mathbf{E} = -\partial \mathbf{B}/\partial t \quad \text{curl } \mathbf{B} = \varepsilon_0 \mu_0 \, \partial \mathbf{E}/\partial t \quad \text{div } \mathbf{E} = \rho/\varepsilon_0 \quad \text{div } \mathbf{B} = 0 \qquad [2A]$$

Maxwell's equations include two very different kinds of electric field, **E**. In the first equation, **E** arises from a changing **B** field and is solenoidal. Solenoidal means that the field has neither sources nor sinks, and so the "lines of force" used to visualize the field are continuous loops. The third equation describes **E** arising from charges, and this field is conservative. The electric force at charge q due to another charge Q, with **E** as intermediary, is described by Coulomb's law:

$$F = (1/4\pi\varepsilon_0)qQ/r^2 \qquad E = (1/4\pi\varepsilon_0)Q/r^2 \qquad \mathbf{F} = q\mathbf{E} \qquad [3A]$$

The **E** field concept was introduced by Faraday. He recognized that "action at a distance" is not logical, and introduced the electric field **E** as the tangible connection between charges. Supposedly sourced by charge, **E** is a conservative field. Conservative means that the work needed to move a charge q from one point to another is independent of the path. This differs from a solenoidal **vxB** electric field, within which a test charge gains (or loses) energy each time it transits a loop of the field. It seemed plausible to Faraday, and has to us ever since, that this conservative **E** field is static, real, carries energy, and arises only from charge. This differs from the **vxB** electric field discussed in this paper, a field that undulates at the Compton frequency and only seems static because current measuring instruments are unresponsive to such rapid oscillation. Maxwell's equations would lead one to believe that **vxB** is always solenoidal, since no charge is involved.

These two kinds of electric field, **E** and **vxB**, are so different that it is surprising that we even continue to use the same symbol, **E**. Lorentz' force law recognizes both, one described as **E** (conservative, from charges) and the other as **vxB** (solenoidal). Until now, it has not been thought that anything other than "charge" could produce a conservative inverse square electric force as measured by a test charge. **vxB** in this model, however, does just that.

It seems impossible that the integral form of Gauss' law finds charge exists within the proposed model of an electron, but div **vxB** cannot find it. How could this be? Gauss' law is not even physics, it is mathematics! It seems incredible that Gauss' law should fail. The answer to the problem lies with whether or not "charge" is a substance. If we were dealing with a real substance, such as water, Gauss' law describes the outflow rate through an enclosing surface as the consequence of sources and sinks of water within that volume. No problem. Makes sense. But since "charge" is not a substance, Gauss' law need not apply. We know of many things (but not substances) for which Gauss' law does not apply, such as love, honor, beauty, etc. It is shocking to realize that Gauss' law does not apply to charge. After all, one of Maxwell's equations is Gauss' law. But, since "charge" is not a substance, Gauss' law is inapplicable and so Maxwell's equations need modification. Charge density may be useful when dealing with many charged particles, just as water can be treated as a fluid instead of a cloud of molecules. But the concept fails utterly at the level of a single particle.

How is it possible that a solenoidal vector field can be arranged to source or sink itself, as suggested in this model of an electron? Sounds like asking a fellow to pull himself up by his bootstraps. Lets look into a mechanism by which the **vxB** field, itself solenoidal and hence divergenceless, can source or sink **vxB** and so produce a conservative electric field. Consider a skinny single loop of **B** field enclosing the blue circle in the next figure. Not to worry about how to produce it, we just assume it and it is so by construction (thanks a lot, theorists). The **B** field orientation is clockwise. When set spinning about a vertical diameter, what do we find? On the left side, **vxB** is directed towards the center and on the right side, **vxB** is directed away from center. What do we call an entity that produces such an electric field? A spinning **electric dipole**. Now move the vertical spin axis to the left, tangent to the loop. Then spin it again and consider the **vxB** field. On the left, it vanishes and on the right it is directed outwards and with twice the magnitude. In the first case, a Gaussian surface drawn about the figure finds zero net electric flux and no charge. The surface integral of **vxB** = 0. In the second case, the Gaussian surface finds a positive definite electric flux and so concludes that a net charge is present within. What do we call the entity that produces such a field? Gauss' law calls it "**charge**". This mechanism, without reliance on any ad hoc charge, can source or sink its own **vxB** electric field and also sink electric fields produced elsewhere. What, then, is charge? Charge is a measure of electrification, assumed incorrectly by Faraday as a substance that "sources" a static inverse square electrical field **E**. This simplified loop of **B** can readily be expanded to form a symmetric magnetic dipole, as in the electron model.

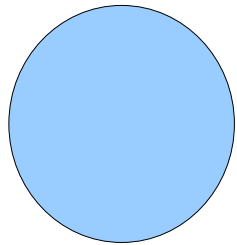

Figure 7. Clockwise loop of B field, initially spinning about a vertical diameter.

In the first case, spinning about a diameter, the **vxB** field at the surface of an enclosing Gaussian sphere is positive on one side and negative on the other and the surface integral is zero. No charge, and so **vxB** is solenoidal. In the second case, the **vxB** field at the surface of an enclosing Gaussian sphere is of one sign only and Gauss' law finds a charge has been enclosed. Charge has been created, and the **vxB** field is now conservative. The distinction between a conservative and a solenoidal electrical field is not so great as has been generally accepted, changing from one to the other with slight changes of geometry within the mechanism that produces the electric field.

If these arguments are persuasive, the basic laws of electromagnetism need to be modified. The static conservative Coulombic **E** field, thought to arise from charge, does not exist. Charge is not a real substance, and only serves as a mathematical convenience when dealing with the electrical forces exerted by one particle on another. The addition of a steady-state offset to the **B** field in this model of an electron is merely a gage transformation. One still needs a reason for stability, and this comes from the spinning **B** fields as mentioned earlier. The basic laws, with apologies for retaining a test charge q in the Lorentz force law, are:

$$\mathbf{F} = q(\mathbf{v \times B}) \qquad [4A]$$

$$\text{curl } \mathbf{v \times B} = -\partial \mathbf{B}/\partial t \qquad \text{curl } \mathbf{B} = \varepsilon_0 \mu_0 \, \partial(\mathbf{v \times B})/\partial t \qquad \langle \text{div } \mathbf{v \times B} \rangle = 0 \qquad \text{div } \mathbf{B} = 0 \qquad [5A]$$

Or, for convenience, we can retain the symbol **E** for the **vxB** electric field but must recognize that this undulating electric field is quite different from Faraday's static **E** field supposedly arising from charges.

A few years ago, I explored the possibility that gravity arises from oscillating long range **E** fields, fields whose range is 1/r. These arise from the first of Maxwell's equations [2A], plus the vector potential defined by **B**=curl A.

$$\mathbf{E} = -\partial \mathbf{A}/\partial t \qquad [6A]$$

The time derivative of **E** has been thought to generate two terms only, but there are really three because "q" in this model undulates at the Compton frequency, ω.

$$q(t) = (q/2)(1 + \sin \omega t) \qquad [7A]$$

$$\mathbf{A} = (\mu_0/4\pi) q \mathbf{v}/r \qquad [8A]$$

$$\mathbf{E} = -\partial \mathbf{A}/\partial t = -(\mu_0/4\pi) q \mathbf{a}/r + (\mu_0/4\pi) q v^2 \mathbf{r}/r^3 - (\mu_0/4\pi)(q \mathbf{v}/2r)\omega \cos\omega t \qquad [9A]$$

It appears that the magnetic field **B** is ultimately the *only* EM field. Charge "q" is merely a convenient mathematical fiction, deriving its attributes from the assumption that such a substance exists that can source or sink the supposedly static inverse square electrical field surrounding certain particles such as electrons and protons. The conservative **vxB** electric field about an electron is not static, but undulates at the Compton frequency, $\nu_C = 1.236 \times 10^{20}$ Hz. Gauss' law, in integral form, remains useful. But in differential form, Gauss' law fails because charge is not a real substance. The EM field and matter as we know it are ultimately comprised of **B** fields plus energy. Nothing else. Of course, it would be foolish to abandon the mathematical crutch of using static conservative **E** fields like those Faraday attributed to charge. It would be foolish to abandon the concept of charge. But when we attempt to really pin down the "why" of things, it is best to winnow out the chaff. The **B** field plus energy appears to constitute the building material from which everything else is formed. This is the "bottom of the barrel".

An answer to "why no magnetic monopoles?" is now clear. So long as there were thought to exist the electrical monopoles known as charges, a deep belief in symmetry demanded that we consider and look for magnetic monopoles. This study concludes that there are no electrical monopoles. To this extent, symmetry has been recovered.